\documentstyle[12pt]{article}
\newcommand{\beq}{\begin{equation}}
\newcommand{\eeq}{\end{equation}}
\newcommand{\bea}{\begin{eqnarray}}
\newcommand{\eea}{\end{eqnarray}}
\def\bar{\begin{array}}
\def\ear{\end{array}}
\def\cap{\par\noindent}
\def\bra{\langle}
\def\ket{\rangle}

\def\le#1{\label{eq:#1}}
\def\re#1{\ref{eq:#1}}

\begin{document}

\noindent
\centerline{\bf Single particle spectrum and binding energy 
 of nuclear matter}

\vskip 0.5 cm

\noindent
\centerline{
M. Baldo$\dag$ and A. Fiasconaro$\ddag$}

\begin{small}

\vskip 0.5 cm
\noindent
\centerline{ $\dag$ INFN, Sez. Catania, 
Corso Italia 57 - 95129 Catania, Italy}

\noindent
\centerline{ $\ddag$ Dipartimento di Fisica, Universita' di Catania,
Corso Italia 57, 95129 Catania, Italy}

\vskip 1 cm

%\hangindent=2.7 true cm \hangafter=0 \noindent
\noindent
{\bf Abstract.} In non-relativistic Brueckner calculations of 
nuclear matter, the self-consistent single particle potential 
is strongly momentum dependent. To simplify the calculations,
a parabolic approximation is often used in the literature.
The variation in the binding energy value introduced by the parabolic
approximation is quantitatively analyzed in detail. It is found that
the approximation can introduce an uncertainty of 1-2 MeV 
near the saturation density.

\vskip 0.6 cm

\noindent{\bf PACS}: 21.65.+f , 24.10.Cn , 45.50.Jf , 21.30.-x , 26.60.+c
\vskip 0.8 cm

\end{small}

\vskip 0.8 cm

\noindent
{\bf 1.~Introduction}
\vskip 0.3 cm
\noindent
It is one of the fundamental issue in nuclear physics to evaluate 
the nuclear matter binding energy and saturation properties, starting
from a realistic nucleon-nucleon (NN) interaction with no free parameter.
This old project requires the solution of a complex many-body problem,
and has received several contributions and improvements along the 
years, beginning as far back as the middle of the last century.
One of the main approaches to this long standing problem is the
so-called hole-expansion or Bethe-Brueckner-Goldstone (BBG) 
theory \cite{book}. The first real breakthrough in this scheme was the 
introduction of the self-consistent single particle potential 
at the two hole-line level of approximation, which is then usually 
referred as the Brueckner-Hartree-Fock (BHF) approximation \cite{gam,book}. 
The introduction of the self-consistent potential drastically improves
the results. In particular, the binding energy and saturation density, 
which otherwise would turn out unreasonable, move to values which can
be considered an acceptable starting approximation. The remaining 
discrepancy could be summarized in the celebrated 
``Coester band" \cite{coest}, along which the saturation points for different
NN interactions were clustering and which misses the empirical region
( corresponding approximately to a binding energy per nucleon of  -16 MeV and
a nucleon density of 0.17 $fm^{-3}$ ).  
Later, the Li\'ege group stressed \cite{mahaux} the relevance of 
the choice of the single particle potential. In particular they suggested
the use of the 
``continuous choice", which indeed appears to move the saturation point
towards the empirical one, but still missing it \cite{mahaux,cont}.
A period of major developments took place in the latest two decades. 
Starting from the works by B.D. Day
\cite{day}, the hole-line expansion was analysed up to the three hole-line
level of approximation. A strong indication of convergence of the expansion
was obtained \cite{day,3hole}. Furthermore, BHF calculations with the
continuous choice seem to get a substantially smaller corrections
from three-body correlations \cite{3hole}. 
The results confirm that the empirical 
saturation point is still missed, and therefore that three-body forces
are needed in the nuclear hamiltonian \cite{day}. In the meanwhile 
the relativistic Dirac-Brueckner (DB) method was developed \cite{dirb}, 
which already at the two hole-line level of approximation appears
to be able to reproduce the empirical saturation point. 
The main relativistic correction introduced by the DB method is due to the
structure of the Dirac 4-spinors, which in the medium appear ``rotated" 
with respect to the free ones. The non-relativistic three-body forces 
and this relativistic effect of the DB
approach are probably two faces of the same dynamical effect \cite{brown}.  
The many-body theory has reached, therefore, such a precision 
that it is possible to test the nuclear hamiltonian. Because of that,
the time seems to be appropriate to check the reliability of the 
approximations which are commonly employed in  BBG calculations
of nuclear matter. \par
In this letter we consider the BHF in the continuous choice and we analyze
quantitatively the uncertainty of the results which comes out
by approximating the single particle self-consistent potential with
a parabolic form. This approximation is quite popular, since it allows
to calculate the potential, at each iteration, only for few momenta, 
thus reducing drastically the computer time. The single particle potential,
as obtained from fully self-consistent BHF calculations, is indeed
strongly momentum dependent and not necessarily so simple as a parabola.
\vskip 0.5 cm
\noindent
{\bf 2.~Sketch of the formalism}
\vskip 0.3 cm
\noindent
In the BHF approximation, the nuclear matter total energy $E$
is obtained from the Brueckner G-matrix $G(\omega)$ according to the equation 
\beq
   E = \sum_{k_1 < k_F} {\hbar^2 k_1^2\over 2m}
 + {1 \over 2} \sum_{k_1,k_2 < k_F}
  \langle k_1 k_2 \vert G(e_{k_1}+e_{k_2}) \vert k_3 k_4 \rangle_A
\le{e2h} 
\eeq
\cap
with $\vert k_1 k_2 \ket_A = \vert k_1 k_2 \ket -
\vert k_2 k_1 \ket $. Here $k_F$ is the Fermi momentum, the summation over
the momenta $k_i$ include spin and isospin variables. The single particle
energies $e_k$, appearing in the entry energy of the G-matrix, are given by
\beq
e(k) = {\hbar^2 k^2\over 2m} + U(k)
\eeq
\cap
where the  single particle potential $U(k)$ is determined
by the self-consistent equation
\beq
 U(k) = \sum_{k' < k_F}
      \langle k k' \vert G(e_{k_1}+e_{k_2}) \vert k  k' \rangle \ \ \ 
\le{auxu}
\eeq
\cap
The self-consistency is coupled with the integral equation for the
G-matrix
\beq
\bar{rl}
 \bra k_1 k_2 \vert G(\omega) \vert k_3 k_4 \ket\!\!\! &\, = \, 
 \bra k_1 k_2 \vert v \vert k_3 k_4 \ket \, +\,  \\
 &                  \\
 + \sum_{k'_3 k'_4} \bra k_1 k_2 \vert v \vert k'_3 k'_4 \ket\!\!\! 
 &{\left(1 - \Theta_F(k'_3)\right) \left(1 - \Theta_F(k'_4)\right)
  \over \omega - e_{k'_3} - e_{k'_4} }
  \, \bra k'_3 k'_4 \vert G(\omega) \vert k_3 k_4 \ket \ \ 
\ear
\le{bruin}
\eeq
\cap
where $\Theta_F(k) = 1$ for $k < k_F$ and is zero otherwise.
The product $Q(k,k') = (1 - \Theta_F(k)) (1 - \Theta_F(k'))$,
appearing in the kernel of  Eq. (\re{bruin}), enforces the scattered
momenta to lie outside the Fermi sphere, and it is commonly referred
as the ``Pauli operator". The self-consistent set of equations are
usually solved by an iteration procedure.
The G-matrix can be expanded in partial waves, according to
the classification of two-nucleon channels \cite{book}.
To avoid coupling between different two-body channels,
the Pauli operator $Q$, as well as the two-body energies
$e_{k'_3} + e_{k'_4}$ in the denominator, are averaged over the
angle between the relative momentum $q = (k'_3 - k'_4)/2$ and the
total momentum $P = k'_3 + k'_4$. Despite this approximation,
which has been tested recently in ref. \cite{muther}, the numerical
solution of the coupled equations (\re{auxu}),(\re{bruin}) is quite time
consuming, since the single particle potential $U(k)$ must be calculated
in a wide range of momenta with a fine enough grid. If one assumes that the
potential $U(k)$, or equivalently the single particle energy $e(k)$, has
approximately a quadratic form
\beq
e(k) \approx e_0 + {\hbar^2 k^2\over 2m^*} 
\le{parab}
\eeq
\cap
then one can calculate the potential, at each iteration step, in few points
only and interpolate the obtained values with a parabola. The approximation
of Eq. (\re{parab}) is usually called the effective mass approximation,
since then the spectrum has the same shape as the free one but with 
an ``effective mass" $m^*$. \par
In order to test this approximation, we have performed a set of BHF
calculations fully self-consistently without any assumption about
the potential shape, as well as by forcing the potential to a 
parabolic shape by means a fitting procedure, and then compared the 
results.
\vskip 0.5 cm
\noindent
{\bf 3.~Results and discussion}
\vskip 0.3 cm
\noindent
The performed BHF calculations include all two-body channels up to total
angular momentum $J = 11\hbar$. In a set of calculations we
adopted the Argonne v$_{18}$ \cite{wir} potential as the NN interaction.
This potential belong to a new generation of realistic NN potentials,
with an improved fit of the scattering data, which give similar
results and cluster closely together in the Coester band \cite{engv}.
The {\it self-consistent} single particle potential $U(k)$ was calculated
up to the momentum cut-off $k_{max} = 7.5 fm^{-1}$, which turns
out to be large enough in the considered density range \cite{future}. 
The potential, for the Fermi momentum $k_F = 1.4 fm^{-1}$, is
displayed in Fig. 1 (full circles). It is numerically calculated with a 
grid step of $0.1 fm^{-1}$ from the G-matrix, Eq. (\re{auxu}), 
and inserted as the entry potential at each iteration step, until
convergence is reached, i.e. the potential and the binding energy 
are stable under iteration with good accuracy. Stability within 
few KeV of the binding was systematically reached.
The numerical method is described in ref. \cite{cont}\par
The quadratic approximation, at each step of 
the iteration procedure, is introduced by fitting the potential
up to a certain maximum momentum $k_{FIT}$. For definiteness, we have
considered in detail two choices, namely $k_{FIT} = 2k_F$ and
$k_{FIT} = k_{max}$. At each iteration step, the potential
$U(k)$ coming directly from the G-matrix calculation
is fitted with a parabola, which is then used as the entry potential for 
the next iteration. Convergence is reached
when both potentials remain stable under this procedure.
In this procedure one obtains, therefore, two potentials,
one calculated from the G-matrix with the {\it parabolic input},
and one from the {\it parabolic fit} to this potential. Of course, if the
the potential coming directly from the G-matrix were indeed parabolic, the 
two potentials would closely agree.
For the choice $k_{FIT} = 2k_F$ , in Fig. 1 the two potentials
at convergence are displayed. In principle, 
one can calculate the nuclear matter binding 
energy from both potentials, but the result will be in general slightly 
different. 
As one can see, the fully self-consistent potential, obtained without any 
fitting procedure, as specified above, does not coincide with anyone of 
the two previous
potentials, and these differences give a quantitative indication of the 
uncertainty introduced by the parabolic approximation. The corresponding 
saturation curves are reported in Fig. 2a. The parabolic potential
produces a saturation curve in fair agreement with the one reported
e.g. in ref. \cite{muther}. Around saturation the parabolic approximation
introduces a shift in the binding of 1-2 MeV. The two choices
for the potentials, discussed above, give different binding, since 
$U(k)$ is not really parabolic, and the fitting
procedure introduces necessarily an approximation. Another
uncertainty is coming from the choice of $k_{FIT}$, as can be
seen in Fig. 2b, where the results for $k_{FIT} = k_{max}$ are reported.
In this case the discrepancy are larger for lower density, since 
then the potential $U(k)$ becomes indeed flatter at 
momenta below $k_F$. A more complete account of the  dependence
on the fitting range is reported in Fig. 3, where the binding
at $k_F = 1.4 fm^{-1}$ is reported as a function of $k_{FIT}$. 
\par
In all cases the saturation curves appear distorted,
and the saturation point shifted. Even if in some cases the 
saturation point seems to be ``improved", this does not have any physical 
meaning, since, anyhow, it is mainly a spurious effect. 
\par
Completely similar results are obtained with the ``old" potential
Argonne v$_{14}$ \cite{wir14}.
\par
In conclusion, we have shown that the parabolic approximation for the 
single particle potential $U(k)$ in the self-consistent Brueckner
scheme introduces an uncertainty of 1-2 MeV near the saturation density,
and therefore it cannot be used in accurate calculations. The full
momentum dependence has to be retained, which prevents the use
of a constant effective mass approximation. However, the uncertainty
is not dramatic, and for approximate estimates of nuclear binding
it can be useful.

\newpage
\noindent
{\bf Figure captions}
\bigskip
\par\noindent
Fig. 1.- Single particle potential as function of momentum. 
 The full circles indicate the results of the fully {\it self-consistent}
 calculation, where the potential is taken at each iteration step
 as calculated from the Brueckner G-matrix. The solid line is the
 result of the parabolic approximation. The parabolic potential,
 used as input for the G-matrix, produces the potential indicated 
 by the squares.
\vskip 0.5 cm 
\par\noindent
Fig. 2.- Saturation curves as obtained for the fully {\it self-consistent}
 calculation (full circles), from the {\it parabolic fit} (triangles)
 and from the single particle potential obtained from the
 {\it parabolic input} (squares). Panel $a$ corresponds to $k_{FIT} = 2 k_F$,
 panel $b$ to $k_{FIT} = k_{max}$.
\vskip 0.5 cm
\par\noindent
Fig. 3.- Dependence of the binding energy from the value of $k_{FIT}$,
 using either the {\it parabolic fit} potential (triangles), or the
 {\it parabolic input} (squares) potential. The arrows indicate
 the value of the binding energy obtained in the fully {\it self-consistent}
 calculation, where no fitting procedure is introduced.

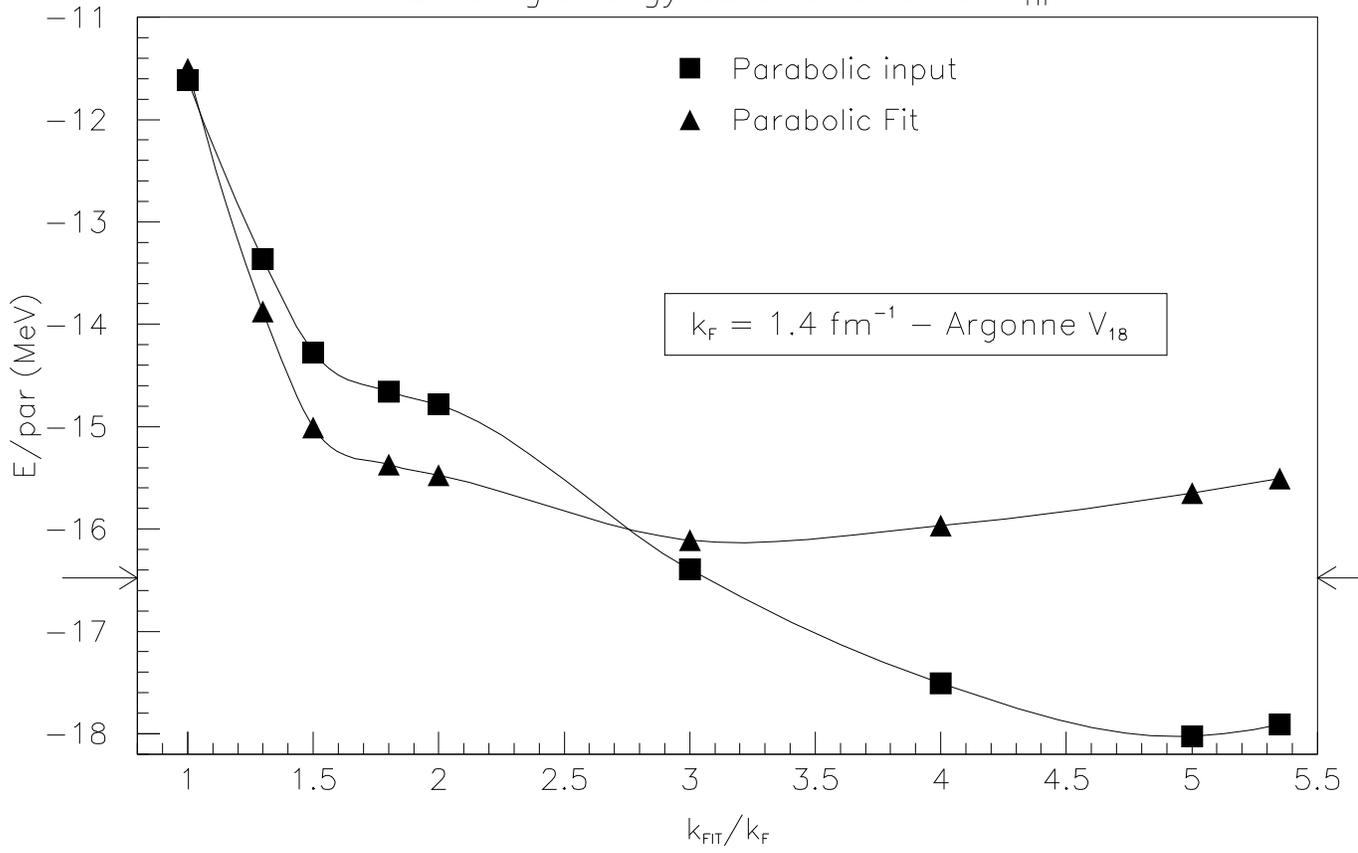
\end{document}